\documentclass{aastex631}
\usepackage[encapsulated]{CJK}
\usepackage{mathrsfs}
\usepackage{subfigure}
\usepackage{epstopdf}
\usepackage{amsmath}
\usepackage{longtable}
\usepackage{booktabs}
\usepackage{hyperref} 
\usepackage{graphicx} 
\usepackage{overpic}
\def\etal {et al.~}

\newbox\grsign \setbox\grsign=\hbox{$>$} \newdimen\grdimen \grdimen=\ht\grsign
\newbox\laxbox \newbox\gaxbox
\setbox\gaxbox=\hbox{\raise.5ex\hbox{$>$}\llap
     {\lower.5ex\hbox{$\sim$}}}\ht1=\grdimen\dp1=0pt
\setbox\laxbox=\hbox{\raise.5ex\hbox{$<$}\llap
     {\lower.5ex\hbox{$\sim$}}}\ht2=\grdimen\dp2=0pt

\shorttitle{Statistics of Spiral Arm Morphology}
\shortauthors{Wei \etal}

\definecolor{malachite}{rgb}{0.34, 0.7, 0.22}

\begin{document}

\title{A New Statistical Analysis of the Morphology of Spiral Galaxies}

\correspondingauthor{Ye Xu}
\email{xuye@pmo.ac.cn}

\author{Junye Wei}
\author{Ye Xu}
\affiliation{Purple Mountain Observatory, Chinese Academy of Sciences, Nanjing 210008, People's Republic of China}
\affiliation{University of Science and Technology of China, 96 Jinzhai Road, Hefei 230026, People's Republic of China}

\author{Zehao Lin}
\author{Chaojie Hao}
\author{Yingjie Li}
\affiliation{Purple Mountain Observatory, Chinese Academy of Sciences, Nanjing 210008, People's Republic of China}

\author{Dejian Liu}
\author{Shuaibo Bian}
\affiliation{Purple Mountain Observatory, Chinese Academy of Sciences, Nanjing 210008, People's Republic of China}
\affiliation{University of Science and Technology of China, 96 Jinzhai Road, Hefei 230026, People's Republic of China}

\begin{abstract}

Morphology is the starting point for understanding galaxies. 
Elmegreen et al. classified spiral galaxies into flocculent, multiple-arm, and grand-design galaxies based on the regularity of their spiral arm structure. 
With the release of a vast number of clear spiral galaxy images from the Sloan Digital Sky Survey, we conducted a morphological classification of 5093 blue spiral galaxies. 
A statistical analysis of this sample shows that the fractions of flocculent, multiple-arm, and grand-design galaxies are 38 $\pm$ 1\%, 59 $\pm$ 1\%, and 3 $\pm$ 1\%, respectively. Redshift has no obvious influence on this classification. However, as the bulge size becomes larger, the fraction of multiple-arm galaxies increases, while that of flocculent galaxies decreases.  
In addition, we performed a statistical analysis of 3958 galaxies with a clear spiral arm structure, finding 82\% of these galaxies have two arms in their inner regions. 
We also found that the majority (74\%) of the barred spiral galaxies exhibit the characteristics of two inner spiral arms and multiple outer spiral arms, and there is no barred spiral galaxy in this work with four continuous spiral arms from the inner to the outer regions. These results highlight that the spiral arm structure of the Milky Way, according to the current mainstream view of a four-arm galaxy with continuous arms extending from the inner to outer regions, is quite unique. However, our findings align with the spiral morphology of the Milky Way proposed by Xu et al., in which case our Galaxy can be considered typical.

\end{abstract}



\section{Introduction}
\label{intro}

A century ago, galaxies were known only as ``nebulae,'' and were found to have a wide range of largely inexplicable forms, with their relationships to one another remaining a mystery. 
It was eventually understood that any theory of galaxy formation and evolution will, at some point, need to account for the vast array of galactic forms \citep{Buta2013}. 
Thus, morphology remains a logical starting point for understanding galaxies. 
Sorting galaxies into their morphological categories is akin to sorting stars into spectral type, which can lead to important astrophysical insights. 

The major survivor of the early visual classification systems is Hubble’s famous ``tuning fork'' \citep{Hubble1926,Hubble1936}, which was later revised and expanded by \cite{Vaucouleurs1959} and \cite{Sandage1961}. 
This system classifies galaxies from ellipticals to spirals. 
Among the trillions of galaxies in the observable universe, spiral galaxies constitute the largest proportion \citep[56\%;][]{Loveday1996}. 
The most prominent feature that sets spiral galaxies apart from other types of galaxies is their spiral arm structure.
The Hubble--de Vaucouleurs classification system discriminates spiral galaxies into subtypes based on several parameters, such as the tightness of the spiral arms \citep{Hubble1926,Hubble1936,Vaucouleurs1959}. 
However, relying solely on this classification system cannot adequately describe the intricate details of spiral arm morphology. 

Based on the regularity of spiral arm structure, \cite{Elmegreen1982} first developed a 12-division morphological classification system, where objects were initially grouped into broad categories: arm classes (AC) 1--4 were designated flocculent galaxies, while AC 5--12 were termed grand-design galaxies. 
Later, AC 10--11 were omitted from the scheme due to the consideration of factors beyond arm morphology, such as obvious stellar bars and companion galaxies \citep{Elmegreen1987}. 
Meanwhile, AC 5--9 were removed from the grand-design category and reclassified as multiple-arm galaxies \citep{Elmegreen1986,Elmegreen1989}. 
In short, based on the regularity of their spiral arm structure, galaxies can be classified as flocculent, multiple arm, or grand design. 
Flocculent galaxies exhibit fragmented and chaotic arms, with poor symmetry and continuity. 
Multiple-arm galaxies feature arms with some symmetry and continuity, often showcasing two symmetric arms in the inner or outer parts. 
Grand-design galaxies possess two long, strongly continuous, and symmetric arms.

By classifying 305 spiral galaxies, \cite{Elmegreen1982} found that multiple-arm and grand-design galaxies account for 69\% of all galaxies in their sample. 
In a subsequent study, \cite{Elmegreen1987} further classified 708 spiral galaxies and realized that among the 654 classifiable spiral galaxies in their sample, flocculent, multiple-arm, and grand-design galaxies constitute proportions of 45\%, 42\%, and 13\%, respectively. 
\cite{Ann2013} classified 1725 spiral galaxies with redshifts $z \leqslant 0.02$ from the Korea Institute for Advanced Study Value Added Galaxy Catalog. 
In their classification results, flocculent, multiple-arm, and grand-design galaxies account for 43\%, 38\%, and 19\% of the total, respectively. 
\cite{Buta2015} classified a sample of 1114 $\rm S^{4}\rm G$ spiral galaxies and found that it contains 50\% flocculent, 32\% multiple-arm, and 18\% grand-design cases.

In addition to classifying spiral arm morphology, \cite{Elmegreen1995} also conducted a statistical analysis of the inner regions (inside about half of each galaxy's radius) of spiral galaxies. 
They obtained a sample of 173 spiral galaxies which possess an inner two-arm spiral structure, where the examined galaxies were obtained from the Atlas of Galaxies \citep{Sandage1988}, the Revised Shapley--Ames Catalog of Bright Galaxies \citep{Sandage1981}, and the Hubble Atlas \citep{Sandage1961}. 
Next, by analyzing the position angles of the end points (where the arms bifurcate or broaden) of the inner arms within each galaxy, \cite{Elmegreen1995} found that the two inner arms of 107 (62\%) galaxies are symmetric to within $20^\circ$.

Currently, the samples used for statistical studies of spiral arm morphology contain around 1000 galaxies, and furthermore, these studies have focused on nearby spiral galaxies with redshifts $z \leqslant 0.02$ \citep{Elmegreen1982,Elmegreen1987,Ann2013}. 
With advancements in observational technology, telescopes have become capable of capturing galaxies at greater distances, enabling us to gather a larger sample of spiral galaxies over a larger redshift range. 
Additionally, the resolution of galaxy images within such samples has greatly improved compared to previous images. 
The increased sample size and enhanced image resolution offer the possibility of more detailed classification statistics regarding the spiral arm structures of spiral galaxies. 
In this work, we compare the results of this paper with the results of \cite{Elmegreen1982,Elmegreen1987,Ann2013,Buta2015}, and analyze whether the proportions of different spiral arm morphologies change under increasing sample size and a wider redshift range. 
Furthermore, the improved image resolution of spiral galaxies allows for more detailed statistics of the inner arm structure, such as the number of inner arms, branching of the spiral arms, the symmetry of arm branching points, and so on. 

Since its start, the Sloan Digital Sky Survey \citep[SDSS;][]{York2000} has completed four phases of sky surveys and is currently conducting its fifth phase \citep{Kollmeier2019}.
This project has detected a large number of galaxies, providing the possibility of conducting more detailed classifications of galaxies.
In order to classify the morphology of the large number of galaxies observed by SDSS, the Galaxy Zoo (GZ1) project was launched in 2007 \citep{Lintott2008, Lintott2011}.
This is a citizen science project, and it provided nearly one million galaxies from SDSS for volunteers to classify.
\cite{Masters2010a} obtained a sample of 5193 blue galaxies from the classification results in GZ1.
Subsequently, the second phase of Galaxy Zoo, Galaxy Zoo 2 (GZ2), selected the best and brightest galaxies from the original GZ1 sample and added some galaxies from SDSS Stripe 82, allowing volunteers to conduct more detailed morphological classifications, including identifying bars, bulges, determining the shapes of edge-on disks, and so on \citep{Willett2013}.
Later, Galaxy Zoo: Hubble (GZH) classified high-redshift galaxies observed with the Hubble Space Telescope \citep[HST;][]{Willett2017}.
The Galaxy Zoo: CANDELS (GZC) has also classified high-redshift galaxies, using HST Advanced Camera for Surveys and Wide Field Camera 3 rest-frame infrared imaging \citep{Simmons2017}.
Additionally, Galaxy Zoo: 3D (GZ:3D) contains a sample of 29,831 galaxies \citep{Masters2021}.
These galaxies were potential targets of the Mapping Nearby Galaxies at Apache Point Observatory survey, part of the fourth phase of SDSS.
In order to facilitate volunteers in drawing the spiral arms of the spiral galaxies in the sample, a subsample that consists of 7418 spiral galaxies with $\leqslant 4$ spiral arms was obtained by \cite{Masters2021}.

In this paper, a statistical study of the spiral arm morphologies of blue spiral galaxies is carried out using SDSS Data Release 18 \citep[DR18;][]{Almeida2023}, with the aim of obtaining statistical data of the spiral arm morphologies of spiral galaxies, and from this, suggesting whether the Milky Way is special in the universe.
Section \ref{sec2} presents the sample used in this study. 
Section \ref{sec3} provides the statistical results of our spiral galaxies, including arm morphology analysis, the numbers of arms of spiral galaxies, the relative positions of branching points for arms with branches, and the symmetry of inner arm branching points. 
Our Milky Way is one of countless spiral galaxies in the universe. 
By leveraging extensive galaxy statistics, we compare the statistical results of this study with existing structural models of the Milky Way in Section \ref{sec4}. 
Conclusions are provided in Section \ref{sec5}.
%

\section{Data}
\label{sec2}

The Galaxy Zoo project provides a large sample of spiral galaxies. 
The original sample of GZ2 is a subset of the sample from GZ1 \citep{Willett2013}.
The galaxies used in GZH and GZC are high-redshift galaxies \citep{Willett2017, Simmons2017}. 
In the sample of 29,831 galaxies from GZ:3D \citep{Masters2021}, there are some elliptical galaxies. 
Moreover, this sample also consists of many edge-on galaxies. 
Its subsample, which consists of 7418 spiral galaxies with $\leqslant 4$ spiral arms may miss most flocculent galaxies. 
Therefore, these samples are not used in this study.

The sample used in this study originates from the work of \cite{Masters2010a}, who selected 5139 blue spiral galaxies from the SDSS Main Galaxy Sample \citep{Strauss2002} by controlling the parameters in GZ1 \citep{Lintott2008} based on SDSS Data Release 6 \citep[DR6;][]{Adelman-McCarthy2008}.
In this sample, the absolute magnitudes of the galaxies are limited to $M_r < -20.17$, and the redshifts are in the range $0.03 < z < 0.085$.
The axial ratios $(a/b)$ from the $r$-band isophotal measurements, which are used as a proxy for disk inclination, are $< 10^{0.2}$, indicating that they are face-on spiral galaxies.
We have updated this sample by using the latest data release from SDSS, i.e., DR18.
Fourteen galaxy coordinates used in GZ1 from SDSS DR6 have become unavailable in SDSS DR18.
Additionally, 32 galaxy images are obstructed by bright objects in the foreground.
After removing these 46 galaxies, a sample containing 5093 blue spiral galaxies was obtained.
Note that since the sample consists of blue spiral galaxies, this sample cannot be representative of all disk galaxies.

\section{Spiral Arm Classification}
\label{sec3}

\subsection{Morphology}

\begin{figure}[!ht]
  \centering
  \subfigure[]{\includegraphics[width=0.6\textwidth]{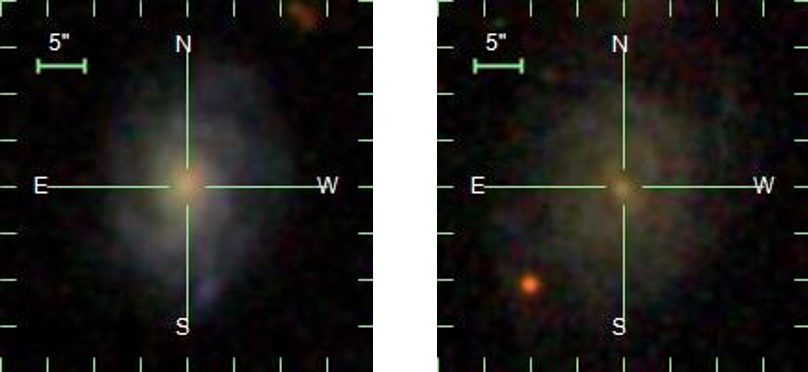}}
  \subfigure[]{\includegraphics[width=0.6\textwidth]{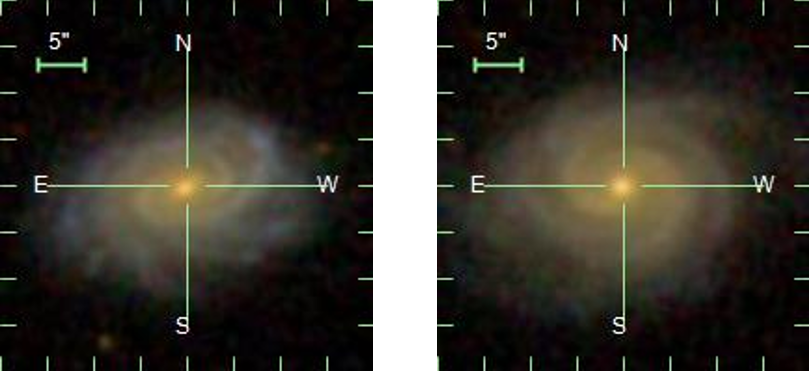}}
  \subfigure[]{\includegraphics[width=0.6\textwidth]{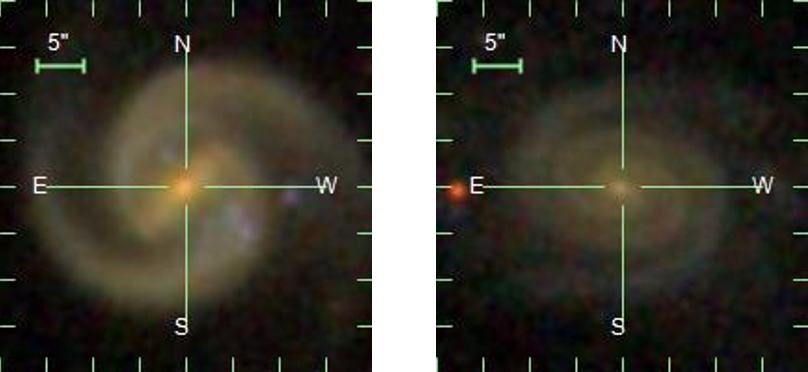}}
  \caption{
    Examples of spiral arms.
    (a) Two flocculent galaxies: SDSS~J122051.70+610237.3 and SDSS~J224158.02-004711.5.
    (b) Two multiple-arm galaxies: SDSS~J151320.27+063704.9 and SDSS~J113223.25+545858.7.
    (c) Two grand-design galaxies: SDSS~J102126.42+381747.3 and SDSS~J092651.03+233037.9.
  }
  \label{figs:morphological}
\end{figure}

Referring to the classification system for spiral arm morphology proposed by \cite{Elmegreen1982,Elmegreen1987}, the spiral arm morphologies of the 5093 galaxies were recorded, as shown in Table \ref{tab1}. 
This table also lists the redshift, the size of each bulge \citep[which is measured by fracdeV, $f_{DeV}$; see][]{Masters2010b}, the existence of a bar, the number of inner arms, the relative positions of the branching points, and the angles of the inner arms' branching points.
This study is based on visual inspections and qualitative impressions of the multi-color images from SDSS. Figure \ref{figs:morphological} shows some typical examples for each type of spiral arm morphology.
In the three morphological types of spiral galaxies, the arms of flocculent galaxies are fragmented, chaotic, and lack almost any symmetry (as shown in Figure \ref{figs:morphological}(a)); grand-design galaxies possess two long and continuous arms with strong symmetry, devoid of branching along the arms (as shown in Figure \ref{figs:morphological}(c)); and multiple-arm galaxies represent an intermediate type, typically featuring symmetric inner or outer arms, with arm continuity and symmetry stronger than in flocculent galaxies but weaker than in grand-design galaxies (as shown in Figure \ref{figs:morphological}(b)). 

\begin{figure}[!ht]
  \centering
  \includegraphics[width=0.7\textwidth]{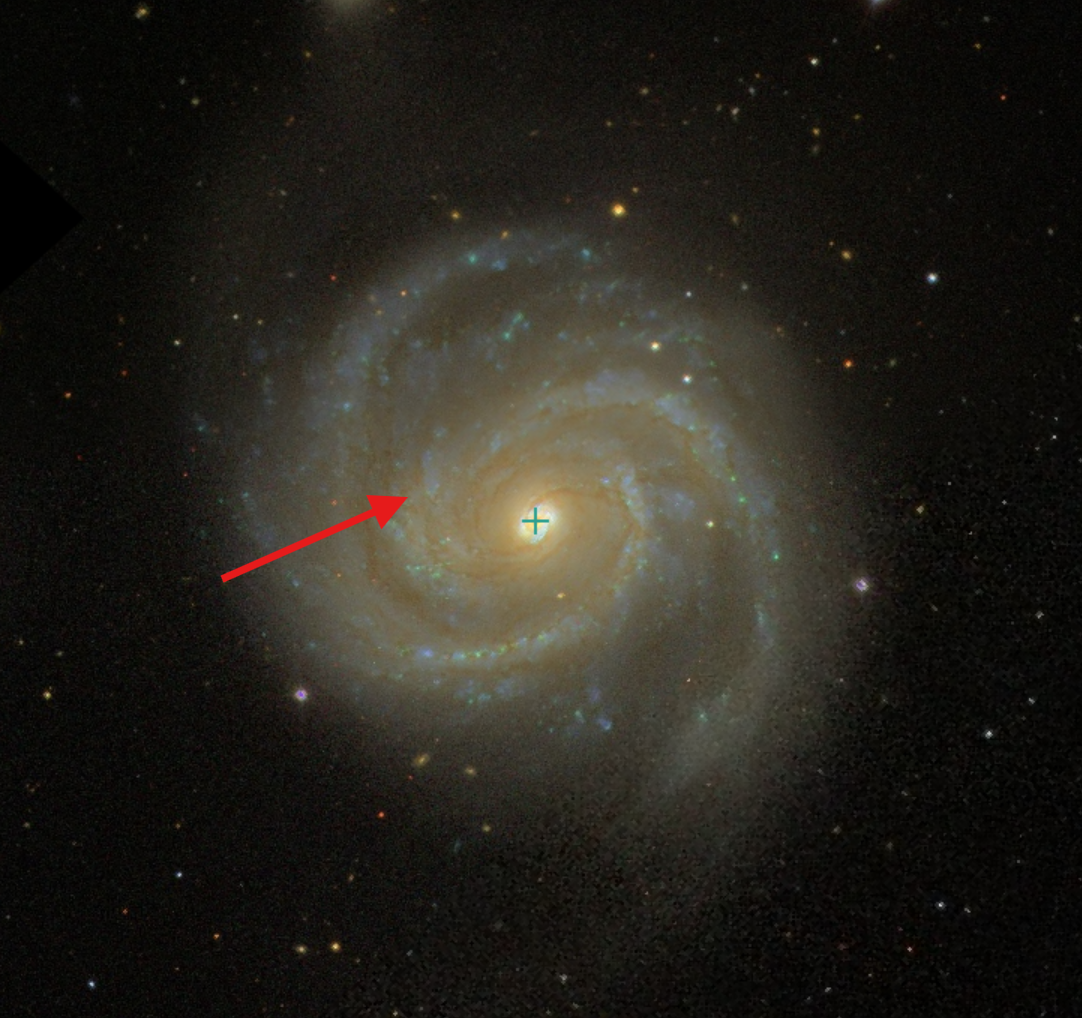}
  \caption{
    NGC~4321. 
    The spiral arm has a clear branch at the position indicated by the arrow. Therefore, it is classified as multiple-arm galaxy. 
  }
  \label{figs:ngc4321}
\end{figure}

We conducted two independent classifications of the sample separated by a one-year interval. In some cases, the second classification of several galaxies is inconsistent with the first classification. However, the difference of proportion for each of the three morphological types obtained from the two classifications is within 2\%.
The results show that the multiple-arm galaxies (59 $\pm$ 1\%) are the most prevalent, followed by flocculent galaxies (38 $\pm$ 1\%) and grand-design galaxies (3 $\pm$ 1\%). Note that the presented values were obtained by the average of the two classifications, and the errors are the difference between the two statistics and the average value.
The fraction of grand-design galaxies is significantly smaller in this study than measured in previous studies (13\% in \citealt{Elmegreen1987}; 19\% in \citealt{Ann2013}; and 18\% in \citealt{Buta2015}), but similar to that ($\sim$ 6\%) found by \cite{Smith2024}.
With the improvement of resolution and sensitivity, some grand-design galaxies also presented bifurcations and/or discontinuities not detected previously, \citep[see][and Figure \ref{figs:ngc4321}]{Xu2023}, meaning several previously classified grand-design galaxies were reclassified as multiple-arm galaxies.

\subsection{Redshift}

\begin{figure}[!ht]
  \centering
  \includegraphics[width=1\textwidth]{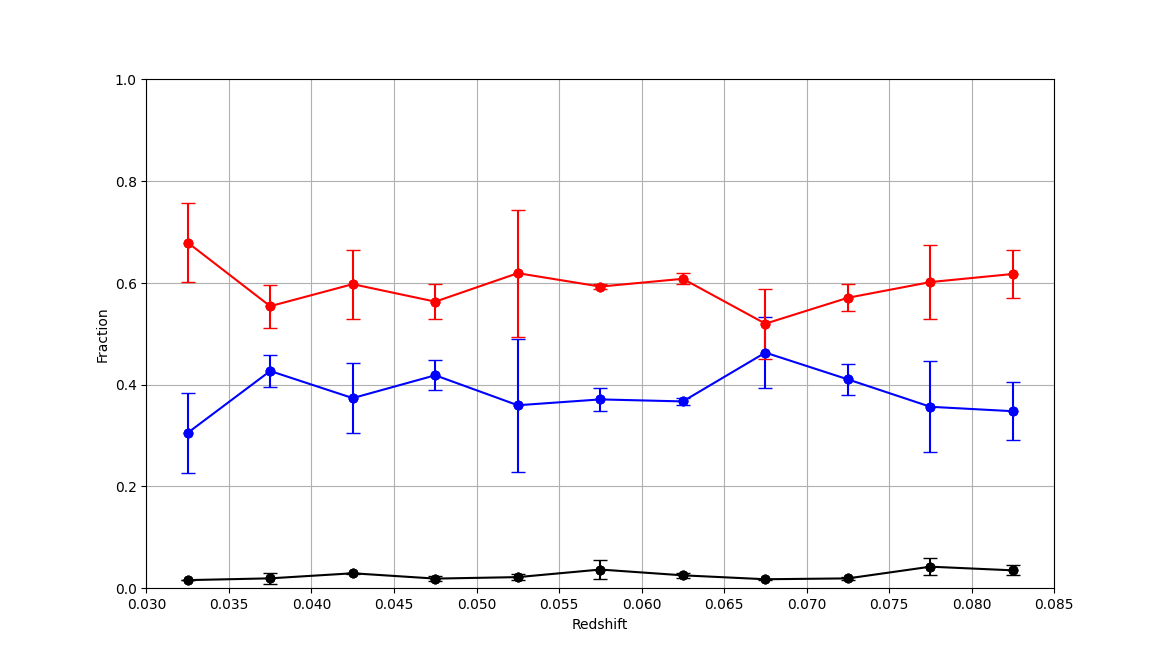}
  \caption{
    Fractions of each morphology in different redshift ranges. 
    Shown are flocculent galaxies (blue), multiple-arm galaxies (red), and grand-design galaxies (black). 
    The errors are presented in the form of error bars.
  }
  \label{figs:redshift}
\end{figure}

The redshifts of our sample range from 0.03 to 0.085, providing favorable conditions for studying whether the proportions of morphological types change as a function of redshift. We subdivided the sample into 11 redshift segments, each spanning 0.005. Figure \ref{figs:redshift} illustrates the relationship between redshift and the morphology fractions. It can be seen that as redshift increases, the fractions of the different spiral arm morphologies show no significant variations.

\subsection{Bulge Size and Bars}

\begin{figure}[!ht]
  \centering
  \includegraphics[width=0.9\textwidth]{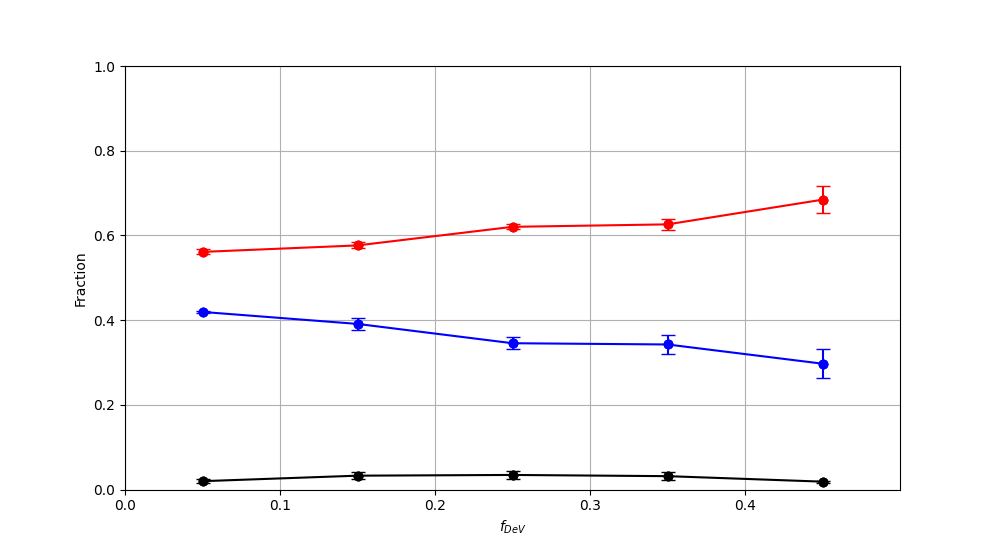}
  \caption{
    Fractions of each morphology in different $f_{DeV}$ ranges. 
    Shown are flocculent galaxies (blue), multiple-arm galaxies (red), and grand-design galaxies (black). 
    The errors are presented in the form of error bars.
  }
  \label{figs:f_DeV}
\end{figure}

The SDSS structural parameter fracdeV, $f_{DeV}$, can indicate the size of a galactic bulge \citep{Masters2010b}.
Spiral galaxies with larger values of $f_{DeV}$ usually have larger bulges.
The $f_{DeV}$ values of the galaxies in our sample are distributed between 0 and 0.5. We divided the $f_{DeV}$ values into five intervals, each with a width of 0.1, and calculated the fraction of each morphology within each interval (see Figure \ref{figs:f_DeV}). We find that as $f_{DeV}$ increases (i.e., the bulge size becomes larger), the fraction of multiple-arm galaxy increases. Conversely, the fraction of flocculent galaxy decreases. 

Another structure in the centers of spiral galaxies, i.e., a bar, was also counted (see Table \ref{tab1}).
We identified the bars visually using the method of \cite{Vaucouleurs1991}.  
The result shows that 29\% of the galaxies in the sample have a bar, while this proportion in flocculent galaxies, multiple-arm galaxies, and grand-design galaxies is 18\%, 36\%, and 37\%, respectively. 
Multiple-arm galaxies and grand-design galaxies have higher fractions of barred spirals than flocculent galaxies. 
The fractions of flocculent, multiple-arm, and grand-design of barred galaxies are 20 $\pm$ 5\%, 77 $\pm$ 4\%, and 3 $\pm$ 1\%, respectively.

\subsection{Spiral Arms}

\begin{figure*}[!ht]
  \centering
  \begin{minipage}[t]{0.4\textwidth}
    \subfigure[]{\includegraphics[width=1\textwidth]{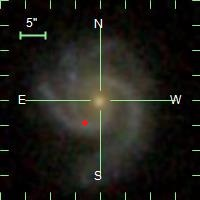}}
  \end{minipage}
  \begin{minipage}[t]{0.4\textwidth}
    \centering
    \subfigure[]{\includegraphics[width=1\textwidth]{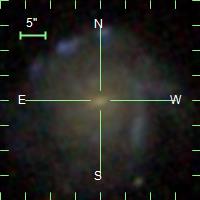}}
  \end{minipage}
  \caption{
    (a) A spiral galaxy with a clear spiral arm structure: SDSS~J151201.76+061309.3.
    Two inner arms can be clearly observed, while a branch of this spiral galaxy can be seen about $5^{\prime\prime}$ southeast of the center (red dot).
    (b) A spiral galaxy with an ambiguous structure: SDSS~J160149.44+173826.0.
    This spiral galaxy has a spiral arm structure, but the spiral arms are too chaotic for us to be able to conduct any further analysis.
  }
  \label{figs:arm_a_d}
\end{figure*}

In our sample, there are 3958 galaxies (Table \ref{tab1}) with a clear structure (see Figure \ref{figs:arm_a_d}(a)) and 1135 galaxies with an ambiguous structure (see Figure \ref{figs:arm_a_d}(b)). Grand-design galaxies exhibit two arms extending from the inner regions all the way to the edges of the galaxies, with no change in the number of arms (see Figure \ref{figs:morphological}(c)).
However, there are differences in the numbers of spiral arms inside and outside of multiple-arm galaxies (see Figure \ref{figs:morphological}(b)) and flocculent galaxies (see Figure \ref{figs:morphological}(a)).
The inner arms of these multiple-arm and flocculent galaxies bifurcate in their disk, leading to this observed difference.
This phenomenon indicates that the number of inner arms and the bifurcation of spiral arms are important parameters to consider when characterizing spiral morphology.

\begin{figure}[!ht]
  \centering
  \subfigure[]{\includegraphics[width=0.35\textwidth]{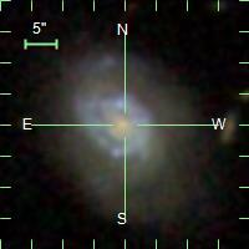}}
  \hspace{2em}
  \subfigure[]{\includegraphics[width=0.35\textwidth]{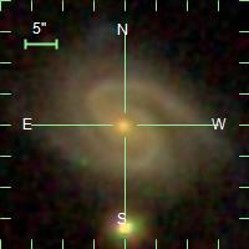}}
  \subfigure[]{\includegraphics[width=0.35\textwidth]{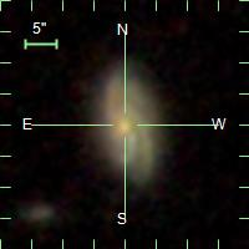}}
  \hspace{2em}
  \subfigure[]{\includegraphics[width=0.35\textwidth]{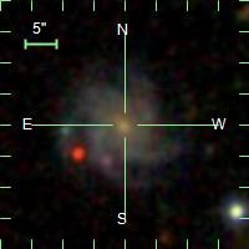}}
  \caption{
    (a) Spiral galaxy with one inner arm, SDSS~J124701.94+403547.1.
    (b) Spiral galaxy with two inner arms, SDSS~J032139.98-063808.2.
    (c) Spiral galaxy with three inner arms, SDSS~J161533.08+262957.0.
    (d) Spiral galaxy with four inner arms, SDSS~J115853.95+014408.3.
  }
  \label{figs:inner_arm}
\end{figure}

Among all 3958 spiral galaxies with a clear structure (including all multiple-arm galaxies, grand-design galaxies, and some flocculent galaxies), the number of inner arms varies.
Table \ref{tab3} lists the number of inner arms for each spiral galaxy type.
There are 3240 (82\%) spiral galaxies with two inner arms, making it the most significant category among galaxies with a clear structure.
Figure \ref{figs:inner_arm} shows some typical examples of galaxies with different numbers of inner arms.

\begin{figure}[!ht]
  \centering
  \includegraphics[width=0.5\textwidth]{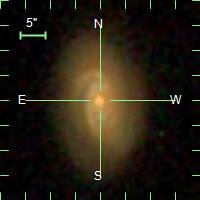}
  \caption{
    This galaxy (SDSS~J230616.43+135856.3) has two arms in the inner region. 
    However, the arms spread out in the outer region.
  }
  \label{figs:spread_out}
\end{figure}

Considering that the Milky Way is a barred spiral galaxy, we also counted the number of barred spiral galaxies with an equal number of inner and outer arms, i.e. $n$ arms that span the inner to the outer regions (see Table \ref{tab5}). 
Note that some galaxies with arms spread out in the outer region are classified as multiple ($>n$) outer arms, such as displayed in Figure \ref{figs:spread_out}.
The majority (74\%) of barred spiral galaxies possess a structure featuring two inner arms and multiple arms in their outer region.
Meanwhile, barred spiral galaxies with four inner arms are extremely rare, and no barred spiral galaxies with four continuous spiral arms from the inner to the outer regions are found in the sample. 

\subsection{Branching Points}

Among the spiral galaxies with a clear structure, 1887 (48\%) have arms with branches, while the remaining 2071 (52\%) have arms without branches or with indistinct branching.
A branching point is the location where a spiral arm branches into multiple spiral arms during its extension. 
We measured the relative positions of the branching points in these galaxies using the following steps.
First, we measure the distance, $R_b$, from the center of the galaxy (from the center of the SDSS image of the galaxy) to the branching point. 
Second, we locate the edge (where the galaxy’s brightness reaches the noise level of the image) on an extension of the line connecting the galactic center and the branching point. Here, $R$ denotes the distance from the edge to the center of the galaxies and the relative position of the branching point is $R_b$/$R$.
This method effectively avoids the impact of the galaxy's inclination along the line of sight on the measurement.

The first statistical result shows that 94\% of the 1887 galaxies with branches have their closest arm branching point located between 0.2 and 0.6 times the galaxy's radius. After a one-year interval, we conducted a second independent check, which yielded a result of 88\%, similar to the first.  Therefore, the relative positions of branching points are concentrated within the 0.2--0.6 $R$ interval , accounting for 91 $\pm$ 3\% of all galaxies with branches.

\begin{figure}[!ht]
  \centering
  \includegraphics[width=0.8\textwidth]{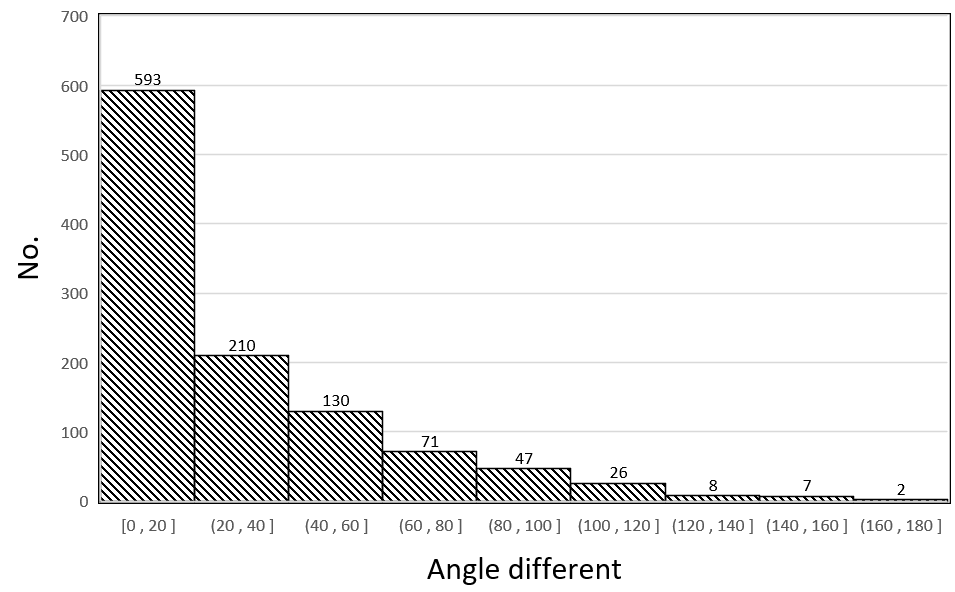}
  \caption{
    Difference in the position angles of the branching points of each arm (with $180^{\circ}$ subtracted) for the multiple-arm galaxies with two inner arms and with both arms branching.
    Many arms are symmetric to within $20^{\circ}$.
  }
  \label{figs:arm_angle}
\end{figure}

\begin{figure}[!ht]
  \centering
  \includegraphics[width=0.7\textwidth]{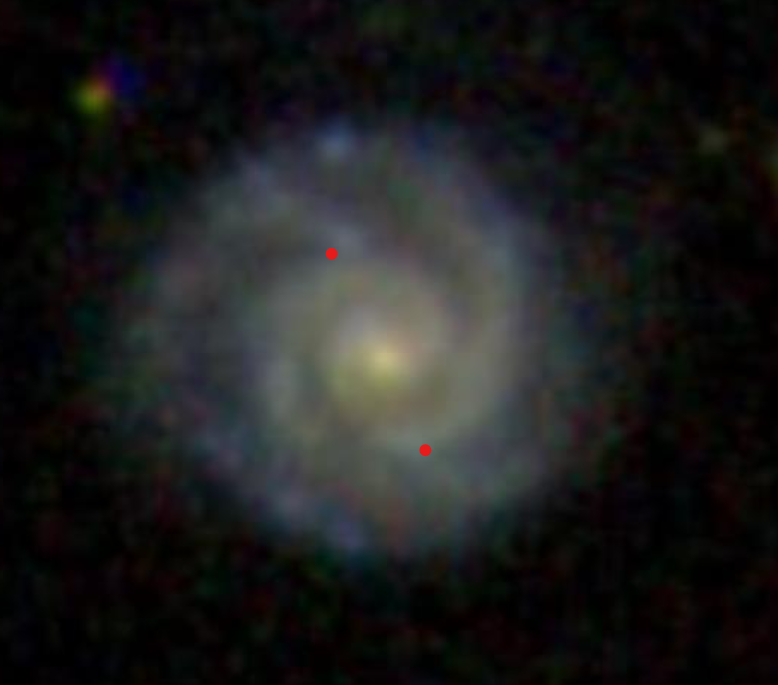}
  \caption{
    A multiple-arm galaxy with an inner two-arm symmetry: SDSS~J113116.03+043908.7. The branching points of its two inner arms (red dots) are symmetric.
  }
  \label{figs:arm_sym}
\end{figure}

A further analysis was made with regard to the 1220 multiple-arm galaxies in our sample that have two inner arms and branching points.
In this sample, there are 1094 multiple-arm galaxies with both arms having branching points, while the others have bifurcation(s) in only one arm (see Figure \ref{figs:arm_a_d}(a) for an example). 
For the former subsample, a histogram (shown in Figure \ref{figs:arm_angle}) of the difference in angle between the branching points of the two arms, rotated by $180^{\circ}$, shows that many arms are within $20^{\circ}$ of being $180^{\circ}$ apart.
Figure \ref{figs:arm_sym} shows an example of a multiple-arm galaxy with an inner two-arm symmetry.
%

\section{Comparing the Results with the Milky Way}
\label{sec4}

Our Milky Way has been considered a spiral galaxy since as far back as the 1850s \citep{Alexander1852}, but it is extremely difficult to observe its spiral structure, due to our edge-on view from its interior and the copious amount of dust extinction along our line of sight. 
It was not until the 1950s that \cite{Morgan1952,Morgan1953} started using high-mass stars (OB stars) to reveal the spiral arm segments in the solar neighborhood. 
Soon after, augmented by radio wavelength data, the larger-scale spiral structure, extending almost across the entire Galactic disk, was mapped using kinematic methods \citep[e.g.,][]{van1954,Oort1958,Bok1959}. 
By analyzing numerous H\textsc{ii}  regions with photometric and/or improved kinematic distances, a paradigmatic map of the Galaxy's spiral arms was created by \citet[][hereafter GG76]{Georgelin1976}, who proposed that the Milky Way has four major arms.
In the subsequent decades of research, significant improvements were made to GG76's four-arm model through the analysis of various spiral-arm tracers, such as molecular clouds \citep[e.g.,][]{Burton1978,Dame1987,Dame2001}, star-forming complexes \citep[e.g.,][]{Russeil2003}, high-mass star-formation regions \citep[e.g.,][]{Reid2019}, and a larger sample of H\textsc{ii} regions \citep[e.g.,][]{Paladini2004,Hou2014}. 
Currently, the mainstream view suggests that the Milky Way is a four-arm galaxy with continuous spiral arms extending from the inner to the outer regions \citep[e.g.,][]{Taylor1993,Drimmel2001,Pettitt2014,Reid2019}. 
However, in our sample, only a few barred spiral galaxies (six out of 1483) have four inner spiral arms. All of these galaxies show that the arms either spread out or bifurcate in the outer region, resulting in more multiple outer arms. This means that none of the barred spiral galaxies exhibit four continuous spiral arms extending from the inner to the outer regions. If the Milky Way does indeed have such a structure, it would be a very unique spiral galaxy in the universe.

Recently, by combining the parallaxes of masers, young open clusters, and massive OB2-type stars, it has been suggested that the Milky Way might possess only two symmetric inner arms and several long irregular outer arms \citep[][hereafter X23]{Xu2023}. 
In the X23 model, the Milky Way is a multiple-arm galaxy with only two inner symmetric arms, which bifurcate as they extend. This morphology is well consistent with most galaxies in our sample as well as the statistical results of this paper.
Specifically, most (74\%, i.e., 1093/1483) barred spiral galaxies exhibit a structure with two inner arms and multiple outer arms (see Table \ref{tab5}). 
Therefore, if the Milky Way is indeed a multiple-arm galaxy with an inner two-arm and an outer multi-arm structure, it would simply be an ordinary spiral galaxy in the universe.
%

\section{Conclusions}
\label{sec5}

In this study, we visually classified 5093 blue spiral galaxies ($0.03 < z < 0.085$), using multi-color images from SDSS DR18.
In our sample, the fractions of flocculent, multiple-arm, and grand-design galaxies are 38 $\pm$ 1\%, 59 $\pm$ 1\%, and 3 $\pm$ 1\%, respectively. Redshift does not have a clear relationship with morphology. Nonetheless, as the bulge size becomes larger, there is a higher fraction of multiple-arm galaxies, while the fraction of flocculent galaxies diminishes.
In the sample, 29\% of the galaxies are barred spiral galaxies, and the bar fractions in grand-design and multiple-arm galaxies are approximately twice that in flocculent galaxies. 
For the spiral galaxies with a clear structure in our sample, the majority (82\%, i.e., 3240/3958) have two inner arms.
Because branching points constitute an important structure of spiral arms, a statistical analysis of the positions of branching points relative to each galaxy's radius was conducted.
The results indicate that the majority of these branching points (91 $\pm$ 3\%) are situated within the range 0.2--0.6 $R$.
Meanwhile, in the multiple-arm galaxies with two inner arms and with both arms branching, many branching points of two inner arms are symmetrical within $20^{\circ}$.

Considering  that the Milky Way is a barred spiral galaxy, we compared its morphology with the statistical results of the barred spiral galaxies in our sample. Only six out of 1483 barred spiral galaxies exhibit four inner spiral arms, and none have been observed with four continuous spiral arms extending from the inner to the outer regions. Actually, the majority (74\%) of barred spiral galaxies display a structure with two inner arms and multiple outer arms. However, the mainstream view suggests that the Milky Way has four arms that continuously extend from the inner to outer regions. Such a distinct structure would make the Milky Way an exceptional case among spiral galaxies. Recently, based on the X23 model, our Galaxy appears more typical, featuring two symmetric inner arms and multiple outer arms, aligning with the structure observed in many other barred spiral galaxies throughout the universe. This interpretation suggests that the Milky Way is not as unique as the four-arm model implies, but rather follows a common galactic pattern.

\clearpage
\startlongtable
\begin{deluxetable}{ccccccccccc}
\setlength\tabcolsep{3pt}
\tablecolumns{11}
\tablecaption{Morphology and Structure of the Spiral Galaxies \label{tab1}}
\tablehead{
\colhead{SDSS ID} & \colhead{Morphology (1)} & \colhead{Morphology (2)} & \colhead{Redshift} & \colhead{$f_{DeV}$} & \colhead{Bar} & \colhead{Inner arm} & \colhead{$R_b/R$ (1)} & \colhead{$R_b/R$ (2)} & \colhead{$\theta_1$} & \colhead{$\theta_2$} \\
 &  &  &  &  & (Y/N) &  &  &  & \colhead{(deg)} & \colhead{(deg)}
}
\startdata
J095603.97+102955.8 & M & M & 0.0300 & 0.34 & N & 2 & 0.32 & 0.38 & 121 & 210 \\
J121252.89+530239.5 & F & F & 0.0300 & 0.40 & N & - &  -   & -    &  -  &  -  \\
J142042.47+095118.1 & F & F & 0.0300 & 0.00 & N & - &  -   & -    &  -  &  -  \\
J084806.89+415133.2 & M & F & 0.0301 & 0.06 & N & 3 & 0.37 & 0.58 &  -  &  -  \\
J092622.42+481000.7 & M & M & 0.0301 & 0.00 & N & 2 & 0.67 & 0.83 &  -  &  -  \\
J172952.39+564934.6 & F & F & 0.0301 & 0.04 & N & - &  -   & -    &  -  &  -  \\
J142644.17+271122.3 & M & M & 0.0301 & 0.43 & Y & 2 & 0.33 & 0.39 & 26  & 81  \\
J164211.73+392719.6 & M & M & 0.0301 & 0.11 & N & 2 & 0.33 & 0.38 & 157 & 355 \\
J170459.90+231008.5 & M & M & 0.0301 & 0.09 & N & 3 &  -   & -    &  -  &  -  \\
J152356.51+380719.7 & F & F & 0.0302 & 0.29 & N & - &  -   & -    &  -  &  -  \\
J145657.58+312922.3 & M & F & 0.0303 & 0.20 & N & 2 & 0.41 & 0.48 & 182 & 342 \\
J122740.41+305553.0 & F & F & 0.0303 & 0.24 & N & - &  -   & -    &  -  &  -  \\
J161403.04+330034.8 & M & M & 0.0303 & 0.00 & Y & 2 & 0.25 & 0.31 &  -  &  -  \\
J224607.46-102209.9 & M & M & 0.0303 & 0.15 & N & 2 & 0.35 & 0.36 &  -  &  -  \\
J122153.55+510233.1 & G & M & 0.0308 & 0.03 & N & 2 &  0   & 0    &  -  &  -  \\
& & &  ...\\
\enddata
\vspace{0.5cm}
\tablecomments{
  Column 1: name of the galaxy.
  Columns 2 and 3: the spiral arm morphologies considered in this work (first and second classifications) are flocculent (F), multiple arm (M), and grand design (G).
  Column 4: redshift of each galaxy.
  Column 5: $f_{DeV}$ of each galaxy.
  Column 6: whether the galactic center has a bar. 
  Column 7: the number of inner arms, with ``-" indicating an uncertain inner arm count.
  Columns 8 and 9: the radius of the galaxy's arm branching points, $R_b$, divided by the galaxy radius, $R$ (first and second classification statistics).
  This parameter takes into account only the branching points closest to the center of the galaxy.
  ``0" indicates that the galaxy has arms without branches or with indistinct branching and ``-" indicates that the arm(s) cannot be counted because they are ambiguous or the arm(s) gradually spread out during their extension, making it impossible to determine the branching point location.
  Columns 10 and 11: the angles of the inner arms' branching points of the multiple-arm galaxies with two inner arms and with both arms branching: $\theta_1$ and $\theta_2$.
  The angles are measured counterclockwise from the western axis.
  (This table is available in its entirety in machine-readable form.)
}
\end{deluxetable}

\clearpage

\startlongtable
\begin{deluxetable}{ccc}
\setlength\tabcolsep{3pt}
\tablecolumns{3}
\tablecaption{Number of Inner Arms \label{tab3}}
\tablehead{
\colhead{No. of inner arms} & \colhead{No. of galaxies} & \colhead{Fraction}
}
\startdata
1 &   43 &  1\% \\
2 & 3240 & 82\% \\
3 &  642 & 16\% \\
4 &   33 &  1\% \\
\enddata
\vspace{0.5cm}
\end{deluxetable}

\startlongtable
\begin{deluxetable}{cccc}
\setlength\tabcolsep{3pt}
\tablecolumns{4}
\tablecaption{Spiral Arms of 1483 Barred Spiral Galaxies \label{tab5}}
\tablehead{
\colhead{Inner arm ($n$)} & \colhead{No. of galaxies} & \colhead{No. of galaxies ($n$ outer arm)} & \colhead{No. of galaxies ($> n$ outer arm)}
}
\startdata
$n$ = 1 &    6 &  0 & 6 \\
$n$ = 2 & 1151 & 58 & 1093 \\
$n$ = 3 &  160 &  5 & 155 \\
$n$ = 4 &    6 &  0 & 6 \\
cannot count &  160 &  - & - \\
\enddata
\vspace{0.5cm}
\tablecomments{Column 1 is the number of inner arms. Column 2 is the number of barred spiral galaxies for each inner arm count. Column 3 is the number of barred spiral galaxies with an equal number of inner and outer arms. Column 4 is the number of barred spiral galaxies whose spiral arms in their outer regions are more numerous than in their inner regions.}
\end{deluxetable}

~\

\begin{acknowledgments}

  We would like to thank the anonymous referee for helpful comments and suggestions that helped to improve the paper. 
  This work was funded by the NSFC grant 11933011, National SKA Program of China (grant No. 2022SKA0120103), and the Key Laboratory for Radio Astronomy.
  This work has made use of data from SDSS-V.
  Funding for the Sloan Digital Sky Survey V has been provided by the Alfred P. Sloan Foundation, the Heising-Simons Foundation, the National Science Foundation, and the Participating Institutions. 
  SDSS acknowledges support and resources from the Center for High-Performance Computing at the University of Utah. 
  SDSS telescopes are located at Apache Point Observatory, funded by the Astrophysical Research Consortium and operated by New Mexico State University, and at Las Campanas Observatory, operated by the Carnegie Institution for Science. 
  The SDSS web site is \url{www.sdss.org}.
  SDSS is managed by the Astrophysical Research Consortium for the Participating Institutions of the SDSS Collaboration, including Caltech, The Carnegie Institution for Science, Chilean National Time Allocation Committee (CNTAC) ratified researchers, The Flatiron Institute, the Gotham Participation Group, Harvard University, Heidelberg University, The Johns Hopkins University, L’Ecole polytechnique f\'{e}d\'{e}rale de Lausanne (EPFL), Leibniz-Institut f\"{u}r Astrophysik Potsdam (AIP), Max-Planck-Institut f\"{u}r Astronomie (MPIA Heidelberg), Max-Planck-Institut f\"{u}r Extraterrestrische Physik (MPE), Nanjing University, National Astronomical Observatories of China (NAOC), New Mexico State University, The Ohio State University, Pennsylvania State University, Smithsonian Astrophysical Observatory, Space Telescope Science Institute (STScI), the Stellar Astrophysics Participation Group, Universidad Nacional Aut\'{o}noma de M\'{e}xico, University of Arizona, University of Colorado Boulder, University of Illinois at Urbana-Champaign, University of Toronto, University of Utah, University of Virginia, Yale University, and Yunnan University.

\end{acknowledgments}

\bibliography{morphology}
\bibliographystyle{aasjournal}

\end{document}